\documentclass[12pt,a4paper]{article}
\makeatletter
\def\eqnarray{%
\stepcounter{equation}%
\let\@currentlabel=\theequation
\global\@eqnswtrue
\global\@eqcnt\z@
\tabskip\@centering
\let\\=\@eqncr
$$\halign to \displaywidth\bgroup\@eqnsel\hskip\@centering
$\displaystyle\tabskip\z@{##}$&\global\@eqcnt\@ne
\hfil$\displaystyle{{}##{}}$\hfil
&\global\@eqcnt\tw@$\displaystyle\tabskip\z@{##}$\hfil
\tabskip\@centering&\llap{##}\tabskip\z@\cr}
\makeatother
\newcommand{\ket}[1]{{\vert{#1}\rangle}}
\newcommand{\bra}[1]{{\langle{#1}\vert}}

\newcommand{\braket}[2]{{\langle{#1}\vert{#2}\rangle}}

\newcommand{\fukuso}{{\mathbf C}}
\newcommand{\real}{{\mathbf R}}
\newcommand{\futon}{{\bf N}}
\newcommand{\seisu}{{\bf Z}}

\newcommand{\zetta}[1]{{\vert{#1}\vert}}

\begin{document}

\title{\sl From Quantum Optics to Non--Commutative Geometry : 
A Non--Commutative Version of the Hopf Bundle, Veronese Mapping and 
Spin Representation}
\author{
  Kazuyuki FUJII
  \thanks{E-mail address : fujii@yokohama-cu.ac.jp}\\
  Department of Mathematical Sciences\\
  Yokohama City University\\
  Yokohama, 236--0027\\
  Japan
  }
\date{}
\maketitle
%
%
%
%
\begin{abstract}
  In this paper we construct a non--commutative version of the Hopf bundle 
  by making use of Jaynes--Commings model and so-called Quantum 
  Diagonalization Method. 
  The bundle has a kind of Dirac strings. However, they appear in only states 
  containing the ground one 
  (${\cal F}\times \{\ket{0}\} \cup \{\ket{0}\}\times {\cal F} \subset 
  {\cal F}\times {\cal F}$) 
  and don't appear in remaining excited states. This means that classical 
  singularities are not universal in the process of non--commutativization. 
  
  Based on this construction we moreover give a non--commutative version of 
  both the Veronese mapping which is the mapping from $\fukuso P^{1}$ to 
  $\fukuso P^{n}$ with mapping degree $n$ and the spin representation of the 
  group $SU(2)$. 
  
  We also present some challenging problems concerning how classical 
  (beautiful) properties can be extended to the non--commutative case. 
\end{abstract}

\newpage

%
%
%
%

\section{Introduction}

This paper is an extended version of \cite{KF}. 

\par \noindent
The Hopf bundles (which are famous examples of fiber bundles) over 
${\bf K}={\bf R}$, ${\bf C}$, ${\bf H}$ (the field of quaternion numbers), 
${\bf O}$ (the field of octanion numbers) are classical objects and 
they are never written down in a local manner. If we write them locally then 
we are forced to encounter singular lines called the Dirac strings, 
see \cite{KF}, \cite{MN}. 

It is very interesting to comment that the Hopf bundles correspond to 
topological solitons called Kink, Monopole, Instanton, Generalized Instanton 
respectively, see for example \cite{MN}, \cite{Ra}, \cite{KFu}. 
Therefore they are very important objects to study in detail. 

Berry has given another expression to the Hopf bundle and Dirac strings by 
making use of a Hamiltonian (a simple spin model including the parameters $x$, 
$y$ and $z$), see the paper(s) in \cite{SW}. 
We call this the Berry model for simplicity. In the following let us restrict 
to the case of ${\bf K}$=${\bf C}$. 

We would like to make the Hopf bundle non--commutative. Whether such a 
generalization is meaningful or not is not clear at the current time, however 
it may be worth trying, see for example \cite{BaI} or more recently 
\cite{SG} and its references.

By the way, we are studying a quantum computation based on Cavity QED and 
one of the basic tools is the Jaynes--Cummings model (or more generally 
the Tavis--Cummings one), \cite{JC}, \cite{MS}, \cite{papers}, \cite{FHKW}. 
This is given as a ``half" of the Dicke model under the resonance condition 
and rotating wave approximation associated to it. 
If the resonance condition is not taken, then this model gives a 
non--commutative version of the Berry model. However, this new one is 
different from usual one because $x$ and $y$ coordinates are quantized, 
while $z$ coordinate is not. 

From the non--commutative Berry model we construct a non--commutative version 
of the Hopf bundle by making use of so--called Quantum Diagonalization Method 
(QDM) developed in \cite{FHKSW}. 
Then we see that the Dirac strings appear in only states containing the ground 
one (${\cal F}\times \{\ket{0}\} \cup \{\ket{0}\}\times {\cal F}$) where 
${\cal F}$ is the Fock space generated by 
$\{a,\ a^{\dagger},\ N=a^{\dagger}a\}$, while they don't appear 
in excited states (${\cal F}\times {\cal F} 
- {\cal F}\times \{\ket{0}\} \cup \{\ket{0}\}\times {\cal F}$).

This means that classical singularities are not universal in the 
process of non--commutativization, which is a very interesting phenomenon. 
This is one of reasons why we consider non--commutative generalizations 
(which are not necessarily unique) of classical geometry.

Moreover, we construct a non--commutative version of the Veronese mapping 
which is the mapping from $\fukuso P^{1}$ to $\fukuso P^{n}$ with mapping 
degree $n$. The mapping degree is usually defined by making use of the 
(first--) Chern class, so our mapping will become important 
if a non--commutative (or quantum) ``Chern class" would be constructed.

We also challenge to construct a non--commutative version of the spin 
representation of group $SU(2)$. However, our trial is not enough 
because we could construct only the cases for spin $j=1$ and $j=3/2$. 
In this problem, we meet a very difficulty arising from the non--commutativity. 
Further study constructing a general theory will be required. 

Why do we consider non--commutative versions of classical field models ? 
What is an advantage to consider such a generalization ? 
Researchers in this subject should answer such natural questions. 
This paper may give one of answers. 

The contents of the paper are as follows :
\begin{flushleft}
{\bf Section 1}\ \ Introduction \\
{\bf Section 2}\ \ Berry Model and Dirac Strings : Review \\
{\bf Section 3}\ \ Non--Commutative Berry Model Arising from the 
Jaynes--Cummings Model \\
{\bf Section 4}\ \ Non--Commutative Hopf Bundle  \\
{\bf Section 5}\ \ Non--Commutative Veronese Mapping \\
{\bf Section 6}\ \ Non--Commutative Representation Theory \\
{\bf Section 7}\ \ Discussion \\
{\bf Appendix}\ \  \\
\quad \quad \quad \ \ {\bf A}\ \ Classical Theory of Projective Spaces \\
\quad \quad \quad \ \ {\bf B}\ \ Local Coordinate of the Projector \\
\quad \quad \quad \ \ {\bf C}\ \ Difficulty of Tensor Decomposition 
\end{flushleft}

\section{Berry Model and Dirac Strings : Review}

First of all we explain the Dirac strings and Hopf bundle which 
Berry constructed in \cite{SW}. 
The Hamiltonian considered by Berry is a simple spin model 
\begin{equation}
\label{eq:berry-hamiltonian}
H_{B}
=x\sigma_{1}+y\sigma_{2}+z\sigma_{3}
=(x-iy)\sigma_{+}+(x+iy)\sigma_{-}+z\sigma_{3}
=
\left(
  \begin{array}{cc}
    z    & x-iy \\
    x+iy & -z
  \end{array}
\right)
\end{equation}
where $\sigma_{j}\ (j=1\sim 3)$ is the Pauli matrices, 
$\sigma_{\pm}\equiv (1/2)(\sigma_{1}\pm i\sigma_{2})$ and $x$, $y$ and $z$ 
are parameters. We would like to diagonalize $H_{B}$ above. 
The eigenvalues are 
\[
\lambda=\pm r\equiv \pm\sqrt{x^{2}+y^{2}+z^{2}}
\]
and corresponding orthonormal eigenvectors are 
\[
\ket{r}=\frac{1}{\sqrt{2r(r+z)}}
 \left(
  \begin{array}{c}
    r+z  \\
    x+iy   
  \end{array}
\right),\quad 
\ket{-r}=\frac{1}{\sqrt{2r(r+z)}}
 \left(
  \begin{array}{c}
    -x+iy  \\
    r+z      
  \end{array}
\right).
\]
Here we assume $(x,y,z) \in \real^{3}-\{(0,0,0)\}\equiv 
\real^{3}\setminus \{0\}$ to avoid a degenerate case. 
Therefore a unitary matrix defined by 
\begin{equation}
U_{I}=(\ket{r},\ket{-r})
=\frac{1}{\sqrt{2r(r+z)}}
\left(
  \begin{array}{cc}
    r+z  & -x+iy \\
    x+iy & r+z
  \end{array}
\right)
\end{equation}
makes $H_{B}$ diagonal like 
\begin{equation}
H_{B}=
U_{I}
\left(
  \begin{array}{cc}
    r &     \\
      & -r
  \end{array}
\right)
U_{I}^{\dagger}\equiv
U_{I}D_{B}U_{I}^{\dagger}.
\end{equation}
We note that the unitary matrix $U_{I}$ is not defined on the whole space 
$\real^{3}\setminus \{0\}$. The defining region of $U_{I}$ is 
\begin{equation}
D_{I}=\real^{3}\setminus \{0\}- \{(0,0,z)\in \real^{3}|\ z < 0\}.
\end{equation}
The removed line $\{(0,0,z)\in \real^{3}|\ z < 0\}$ 
is just the (lower) Dirac string, which is impossible to add to $D_{I}$.

Next, we have another diagonal form of $H_{B}$ like
\begin{equation}
H_{B}=U_{II}D_{B}U_{II}^{\dagger}
\end{equation}
with the unitary matrix $U_{II}$ defined by 
\begin{equation}
U_{II}
=\frac{1}{\sqrt{2r(r-z)}}
\left(
  \begin{array}{cc}
    x-iy & -r+z \\
    r-z  & x+iy
  \end{array}
\right).
\end{equation}
The defining region of $U_{II}$ is 
\begin{equation}
D_{II}=\real^{3}\setminus \{0\}- \{(0,0,z)\in \real^{3}|\ z > 0\}.
\end{equation}
The removed line $\{(0,0,z)\in \real^{3}|\ z > 0\}$ 
is the (upper) Dirac string, which is also impossible to add to $D_{II}$.

\begin{figure}
\begin{center}
\setlength{\unitlength}{1mm} 
\begin{picture}(140,60)(5,0)
\put(40,20){\circle{1}}
\put(20,20){\line(1,0){19}}
\put(41,20){\vector(1,0){19}}
\put(40,21){\vector(0,1){19}}
\multiput(40,19)(0,-2.5){8}{\line(0,-1){2}}
\put(40.4,20.4){\line(1,1){15}}
\put(39.7,19.7){\vector(-1,-1){15}}
\put(100,20){\circle{1}}
\put(80,20){\line(1,0){19}}
\put(101,20){\vector(1,0){19}}
\multiput(100,21)(0,2.5){7}{\line(0,1){2}}
\put(100,37.5){\vector(0,1){2}}
\put(100,19){\line(0,-1){19}}
\put(100.4,20.4){\line(1,1){15}}
\put(99.7,19.7){\vector(-1,-1){15}}
\end{picture}
\end{center}
\caption{Dirac strings corresponding to I and II}
\end{figure}
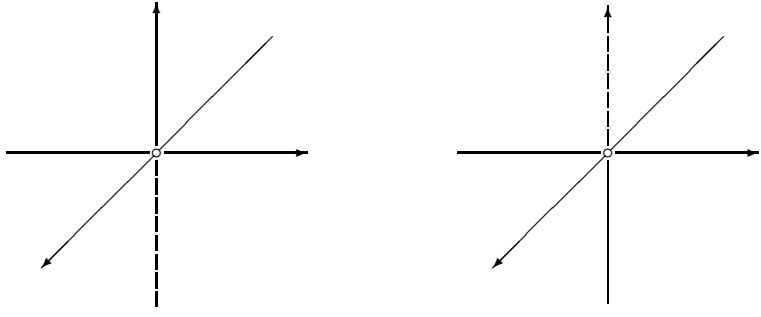

Here we have diagonalizations of two types for $H_{B}$, so a natural question 
comes about. What is a relation between $U_{I}$ and $U_{II}$ ? 
If we define 
\begin{equation}
\Phi
=\frac{1}{\sqrt{x^{2}+y^{2}}}
\left(
  \begin{array}{cc}
    x-iy &      \\
         & x+iy
  \end{array}
\right)
\end{equation}
then it is easy to see 
\[
U_{II}=U_{I}\Phi.
\]
We note that $\Phi$ (which is called a transition function) is not defined on 
the whole $z$--axis. 

What we would like to emphasize here is that the diagonalization of $H_{B}$ is 
not given globally (on $\real^{3}\setminus \{0\}$). However, the dynamics is 
perfectly controlled by the system 
\begin{equation}
\label{eq:Hopf-bundle}
\left\{(U_{I},D_{I}), (U_{II},D_{II}), \Phi, 
D_{I}\cup D_{II}=\real^{3}\setminus \{0\}\right\},
\end{equation}
which defines a famous fiber bundle called the Hopf bundle associated to the 
complex numbers ${\bf C}$ \footnote{The base space $\real^{3}\setminus \{0\}$ 
is homotopic to the two--dimensional sphere $S^{2}$},
\[
S^{1}\longrightarrow S^{3}\longrightarrow S^{2},
\]
see \cite{MN}.

The projector corresponding to the Hopf bundle is given as 
\begin{equation}
\label{eq:projector}
P(x,y,z)=U_{I}P_{0}U_{I}^{\dagger}=U_{II}P_{0}U_{II}^{\dagger}
=
\frac{1}{2r}
\left(
  \begin{array}{cc}
    r+z  & x-iy  \\
    x+iy & r-z
  \end{array}
\right),
\end{equation}
where $P_{0}$ is a basic one 
\[
P_{0}=
\left(
  \begin{array}{cc}
    1 &   \\
      & 0
  \end{array}
\right)\ \ {\in}
 \ \ M(2,{\bf C}).
\]
We note that in (\ref{eq:projector}) Dirac strings don't appear because 
the projector $P$ is expressed globally.

\section{Non--Commutative Berry Model Arising from the \\
Jaynes--Cummings Model}

First, let us explain the Jaynes--Cummings model which is well--known in 
quantum optics, \cite{JC}, \cite{MS}. 
The Hamiltonian of Jaynes--Cummings model can be written as follows (we set 
$\hbar=1$ for simplicity) 
\begin{equation}
\label{eq:hamiltonian}
H=
\omega 1_{2}\otimes a^{\dagger}a + \frac{\Delta}{2} \sigma_{3}\otimes {\bf 1} 
+ g\left(\sigma_{+}\otimes a+\sigma_{-}\otimes a^{\dagger} \right),
\end{equation}
where $\omega$ is the frequency of single radiation field, $\Delta$ the energy 
difference of two level atom, $a$ and $a^{\dagger}$ are annihilation and 
creation operators of the field, and $g$ a coupling constant. We assume that 
$g$ is small enough (a weak coupling regime). 
Here $\sigma_{+}$, $\sigma_{-}$ and $\sigma_{3}$ are given as 
\begin{equation}
\label{eq:sigmas}
\sigma_{+}=
\left(
  \begin{array}{cc}
    0& 1 \\
    0& 0
  \end{array}
\right), \quad 
\sigma_{-}=
\left(
  \begin{array}{cc}
    0& 0 \\
    1& 0
  \end{array}
\right), \quad 
\sigma_{3}=
\left(
  \begin{array}{cc}
    1& 0  \\
    0& -1
  \end{array}
\right), \quad 
1_{2}=
\left(
  \begin{array}{cc}
    1& 0  \\
    0& 1
  \end{array}
\right).
\end{equation}
See the figure 2 as an image of the Jaynes--Cummings model (we don't repeat 
here). 

\begin{figure}
\begin{center}
\setlength{\unitlength}{1mm} 
\begin{picture}(80,40)(0,-20)
\bezier{200}(20,0)(10,10)(20,20)
\put(20,0){\line(0,1){20}}
\put(30,10){\circle*{3}}
\bezier{200}(40,0)(50,10)(40,20)
\put(40,0){\line(0,1){20}}
\put(10,10){\dashbox(40,0)}
\put(49,9){$>$}
\put(54,9){a photon}
\end{picture}
\end{center}
\vspace{-20mm}
\caption{One atom and a single photon inserted in a cavity}
\end{figure}
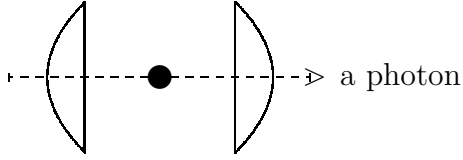

Now we consider the evolution operator of the model. 
We rewrite the Hamiltonian (\ref{eq:hamiltonian}) as follows. 
\begin{equation}
\label{eq:hamiltonian-rewrite}
H=
\omega 1_{2}\otimes a^{\dagger}a + \frac{\omega}{2}\sigma_{3}\otimes {\bf 1} + 
\frac{\Delta-\omega}{2} \sigma_{3}\otimes {\bf 1} +
g\left(\sigma_{+}\otimes a+\sigma_{-}\otimes a^{\dagger} \right)
\equiv H_{1}+H_{2}.
\end{equation}
Then it is easy to see $[H_{1},H_{2}]=0$, which leads to 
$
\mbox{e}^{-itH}=\mbox{e}^{-itH_{1}}\mbox{e}^{-itH_{2}}.
$

In the following we consider $\mbox{e}^{-itH_{2}}$ in which 
the resonance condition $\Delta-\omega=0$ is not taken. For simplicity 
we set $\theta=\frac{\Delta-\omega}{2g}(\neq 0)$ \footnote{Since the 
Jaynes--Cummings model is obtained by the Dicke model under some resonance 
condition on parameters included, it is nothing but an approximate one 
in the neighborhood of the point, so we must assume that $|\theta|$ is small 
enough. However, as a model in mathematical physics there is no problem 
to take $\theta$ be arbitrary} then 
\[
H_{2}
=g\left(
\sigma_{+}\otimes a+\sigma_{-}\otimes a^{\dagger} + 
\frac{\Delta-\omega}{2g}\sigma_{3}\otimes {\bf 1}
\right)
=g\left(
\sigma_{+}\otimes a+\sigma_{-}\otimes a^{\dagger} + 
\theta \sigma_{3}\otimes {\bf 1}
\right).
\]
For further simplicity we set 
\begin{equation}
\label{eq:non-commutative Berry model}
H_{JC}
=\sigma_{+}\otimes a+\sigma_{-}\otimes a^{\dagger} + 
\theta \sigma_{3}\otimes {\bf 1}
=
\left(
  \begin{array}{cc}
    \theta      & a       \\
    a^{\dagger} & -\theta
  \end{array}
\right),\quad [a,a^{\dagger}]={\bf 1}
\end{equation}
where we have written $\theta$ in place of $\theta {\bf 1}$ for simplicity. 

$H_{JC}$ can be considered as a non-commutative version of $H_{B}$ under the 
quantum-classical correspondence 
$
a\ \longleftrightarrow\ x-iy,\ a^{\dagger}\ \longleftrightarrow\ x+iy \ 
\mbox{and}\ \theta\ \longleftrightarrow\ z
$ : 
\[
H_{B}
=
\left(
  \begin{array}{cc}
    z    & x-iy \\
    x+iy & -z
  \end{array}
\right),\ [x-iy,x+iy]=0 \quad \longrightarrow \quad
H_{JC}
=
\left(
  \begin{array}{cc}
    \theta      & a       \\
    a^{\dagger} & -\theta
  \end{array}
\right),\ [a,a^{\dagger}]={\bf 1}.
\]
That is, $x$ and $y$ coordinates are quantized, while $z$ coordinate is not, 
which is different from usual one, see for example \cite{BaI}. 
It may be possible for us to call this {\bf a non--commutative Berry model}. 
We note that this model is derived not ``by hand" but by the model in quantum 
optics itself.

\section{Non--Commutative Hopf Bundle}

We usually analyze (\ref{eq:non-commutative Berry model}) by reducing it to 
each component contained in $H(2,{\bf C})$, which is a typical analytic 
method. However, we don't adopt such a method. 

First we make the Hamiltonian (\ref{eq:non-commutative Berry model}) 
diagonal like in Section 2 and research whether ``Dirac strings" exist or not 
in this non--commutative model, which is very interesting from not only 
quantum optical but also mathematical point of view.

It is easy to see 
\begin{equation}
\label{eq:non-commutative decomposition}
H_{JC}
=
\left(
  \begin{array}{cc}
    \theta      & a       \\
    a^{\dagger} & -\theta
  \end{array}
\right)
=
\left(
  \begin{array}{cc}
    1 &                                 \\
      & a^{\dagger}\frac{1}{\sqrt{N+1}}
  \end{array}
\right)
\left(
  \begin{array}{cc}
     \theta   & \sqrt{N+1} \\
   \sqrt{N+1} & -\theta
  \end{array}
\right)
\left(
  \begin{array}{cc}
    1 &                       \\
      & \frac{1}{\sqrt{N+1}}a
  \end{array}
\right)
\end{equation}
from \cite{FHKSW}, where $N$ is the number operator $N=a^{\dagger}a$. 
Then the middle matrix in the right hand side can be considered as a classical 
one, so we can diagonalize it easily 
\begin{equation}
\label{eq:middle matrix decomposition}
\left(
  \begin{array}{cc}
     \theta   & \sqrt{N+1} \\
   \sqrt{N+1} & -\theta
  \end{array}
\right)
=
\left\{
\begin{array}{l}
 U_{I}
 \left(
  \begin{array}{cc}
    R(N+1) &         \\
           & -R(N+1)
  \end{array}
 \right)
 U_{I}^{\dagger}\\
 U_{II}
 \left(
  \begin{array}{cc}
    R(N+1) &         \\
           & -R(N+1)
  \end{array}
 \right)
 U_{II}^{\dagger} 
\end{array}
\right.
\end{equation}
where 
\[
R(N)=\sqrt{N+\theta^{2}} 
\]
and $U_{I}$, $U_{II}$ are given by 
\begin{eqnarray}
\label{eq:unitary-i}
U_{I}&=&
\frac{1}{\sqrt{2R(N+1)(R(N+1)+\theta)}}
\left(
  \begin{array}{cc}
   R(N+1)+\theta & -\sqrt{N+1}   \\
   \sqrt{N+1}    & R(N+1)+\theta
  \end{array}
\right),  \\
\label{eq:unitary-ii}
U_{II}&=&
\frac{1}{\sqrt{2R(N+1)(R(N+1)-\theta)}}
\left(
  \begin{array}{cc}
      \sqrt{N+1}   & -R(N+1)+\theta \\
     R(N+1)-\theta & \sqrt{N+1} 
  \end{array}
\right). 
\end{eqnarray}
Now let us rewrite (\ref{eq:non-commutative decomposition}) by making use of 
(\ref{eq:middle matrix decomposition}) with (\ref{eq:unitary-i}). 
Inserting the identity 
\[
\left(
  \begin{array}{cc}
    1 &                       \\
      & \frac{1}{\sqrt{N+1}}a
  \end{array}
\right)
\left(
  \begin{array}{cc}
    1 &                                 \\
      & a^{\dagger}\frac{1}{\sqrt{N+1}}
  \end{array}
\right)
=
\left(
  \begin{array}{cc}
    1 &   \\
      & 1
  \end{array}
\right)
\]
gives 
\begin{eqnarray}
H_{JC}=
&&\left(
  \begin{array}{cc}
    1 &                                 \\
      & a^{\dagger}\frac{1}{\sqrt{N+1}}
  \end{array}
\right)
U_{I}
\left(
  \begin{array}{cc}
    R(N+1) &         \\
           & -R(N+1)
  \end{array}
\right)
U_{I}^{\dagger}
\left(
  \begin{array}{cc}
    1 &                       \\
      & \frac{1}{\sqrt{N+1}}a
  \end{array}
\right)    \nonumber    \\
=
&&
\left(
  \begin{array}{cc}
    1 &                                 \\
      & a^{\dagger}\frac{1}{\sqrt{N+1}}
  \end{array}
\right)
U_{I}
\left(
  \begin{array}{cc}
    1 &                       \\
      & \frac{1}{\sqrt{N+1}}a
  \end{array}
\right)
\left(
  \begin{array}{cc}
    1 &                                 \\
      & a^{\dagger}\frac{1}{\sqrt{N+1}}
  \end{array}
\right)
\left(
  \begin{array}{cc}
    R(N+1) &         \\
           & -R(N+1)
  \end{array}
\right)\times   \nonumber  \\
&&
\left(
  \begin{array}{cc}
    1 &                       \\
      & \frac{1}{\sqrt{N+1}}a
  \end{array}
\right)
\left(
  \begin{array}{cc}
    1 &                                 \\
      & a^{\dagger}\frac{1}{\sqrt{N+1}}
  \end{array}
\right)
U_{I}^{\dagger}
\left(
  \begin{array}{cc}
    1 &                       \\
      & \frac{1}{\sqrt{N+1}}a
  \end{array}
\right)         \nonumber  \\
=
&&
V_{I}
\left(
  \begin{array}{cc}
    R(N+1) &         \\
           & -R(N)
  \end{array}
\right)
V_{I}^{\dagger}, 
\end{eqnarray}
where
\begin{eqnarray}
\label{eq:V-I}
V_{I}
&=&
\left(
  \begin{array}{cc}
    \frac{1}{\sqrt{2R(N+1)(R(N+1)+\theta)}} &      \\
           & \frac{1}{\sqrt{2R(N)(R(N)+\theta)}}
  \end{array}
\right)
\left(
  \begin{array}{cc}
      R(N+1)+\theta & -a            \\
      a^{\dagger}   &  R(N)+\theta
  \end{array}
\right)         \nonumber \\
&=&
\left(
  \begin{array}{cc}
      R(N+1)+\theta & -a            \\
      a^{\dagger}   &  R(N)+\theta
  \end{array}
\right)  
\left(
  \begin{array}{cc}
    \frac{1}{\sqrt{2R(N+1)(R(N+1)+\theta)}} &      \\
           & \frac{1}{\sqrt{2R(N)(R(N)+\theta)}}
  \end{array}
\right).
\end{eqnarray}

Similarly, we can rewrite (\ref{eq:non-commutative decomposition}) by making 
use of (\ref{eq:middle matrix decomposition}) with (\ref{eq:unitary-ii}). 
By inserting the identity 
\[
\left(
  \begin{array}{cc}
    \frac{1}{\sqrt{N+1}}a &    \\
                          & 1
  \end{array}
\right)
\left(
  \begin{array}{cc}
    a^{\dagger}\frac{1}{\sqrt{N+1}} &    \\
                                    & 1
  \end{array}
\right)
=
\left(
  \begin{array}{cc}
    1 &   \\
      & 1
  \end{array}
\right)
\]
we obtain 
\begin{equation}
H_{JC}=
V_{II}
\left(
  \begin{array}{cc}
    R(N) &          \\
         & -R(N+1)
  \end{array}
\right)
V_{II}^{\dagger}, 
\end{equation}
where 
\begin{eqnarray}
\label{eq:V-II}
V_{II}
&=&
\left(
  \begin{array}{cc}
    \frac{1}{\sqrt{2R(N+1)(R(N+1)-\theta)}} &     \\
           & \frac{1}{\sqrt{2R(N)(R(N)-\theta)}}
  \end{array}
\right)
\left(
  \begin{array}{cc}
          a       & -R(N+1)+\theta  \\
      R(N)-\theta &  a^{\dagger} 
  \end{array}
\right)        \nonumber \\
&=&
\left(
  \begin{array}{cc}
          a       & -R(N+1)+\theta  \\
      R(N)-\theta &  a^{\dagger} 
  \end{array}
\right) 
\left(
  \begin{array}{cc}
    \frac{1}{\sqrt{2R(N)(R(N)-\theta)}} &             \\
           & \frac{1}{\sqrt{2R(N+1)(R(N+1)-\theta)}}
  \end{array}
\right).
\end{eqnarray}

Tidying up these we have 
\begin{equation}
\label{eq:final-decomposition}
H_{JC}=
\left\{
\begin{array}{l}
V_{I}
\left(
  \begin{array}{cc}
    R(N+1) &         \\
           & -R(N)
  \end{array}
\right)
V_{I}^{\dagger}\\
V_{II}
\left(
  \begin{array}{cc}
    R(N) &         \\
         & -R(N+1)
  \end{array}
\right)
V_{II}^{\dagger}
\end{array}
\right.
\end{equation}
with $V_{I}$ and $V_{II}$ above. From the equations 
\[
R(N+1)\ket{0}=\sqrt{1+\theta^{2}}>\theta,\quad
R(N)\ket{0}=\sqrt{\theta^{2}}=|\theta|
\]
we know 
\[
\left(R(N)\pm \theta\right)\ket{0}=\left(|\theta|\pm \theta\right)\ket{0},
\]
so the strings corresponding to Dirac ones exist in only states 
${\cal F}\times \{\ket{0}\} \cup \{\ket{0}\}\times {\cal F}$ where ${\cal F}$ 
is the Fock space generated by $\{a,\ a^{\dagger},\ N\}$, while in other 
excited states ${\cal F}\times {\cal F}\setminus 
{\cal F}\times \{\ket{0}\} \cup \{\ket{0}\}\times {\cal F}$ they don't exist
\footnote{We have identified ${\cal F}\times {\cal F}$ with the space of 
$2$--component vectors over ${\cal F}$}, see the figure 3. 
The phenomenon is very interesting. 
For simplicity we again call these strings Dirac ones in the following. 

The ``parameter space" of $H_{JC}$ can be identified with 
${\cal F}\times {\cal F}\times \real \ni (*, *, \theta)$, so the domains 
$D_{I}$ of $V_{I}$ and $D_{II}$ of $V_{II}$ are respectively 
\begin{eqnarray}
D_{I}&=&{\cal F}\times {\cal F}\times \real - 
{\cal F}\times \{\ket{0}\}\times \real_{\leq 0},  \\
D_{II}&=&{\cal F}\times {\cal F}\times \real - 
\left({\cal F}\times \{\ket{0}\} \cup \{\ket{0}\}\times {\cal F}\right)
\times \real_{\geq 0}
\end{eqnarray}
by (\ref{eq:V-I}) and (\ref{eq:V-II}). We note that 
\[
D_{I}\cup D_{II}={\cal F}\times {\cal F}\times \real - 
{\cal F}\times \{\ket{0}\}\times \{\theta=0\}.
\]
%

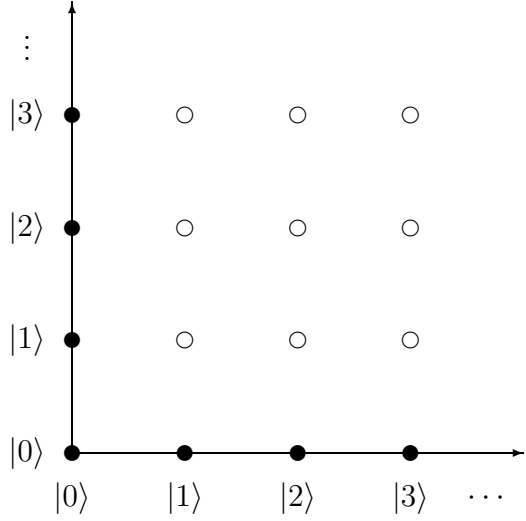
\begin{figure}
\begin{center}
\setlength{\unitlength}{1mm} 
\begin{picture}(80,70)
\put(10,10){\vector(1,0){60}}
\put(10,10){\vector(0,1){60}}
\put(10,10){\circle*{2}}
\put(25,10){\circle*{2}}
\put(40,10){\circle*{2}}
\put(55,10){\circle*{2}}
%
\put(10,25){\circle*{2}}
\put(10,40){\circle*{2}}
\put(10,55){\circle*{2}}
\put(25,25){\circle{2}}
\put(25,40){\circle{2}}
\put(25,55){\circle{2}}
\put(40,25){\circle{2}}
\put(40,40){\circle{2}}
\put(40,55){\circle{2}}
\put(55,25){\circle{2}}
\put(55,40){\circle{2}}
\put(55,55){\circle{2}}
\put(7,1){\makebox(6,6)[c]{$|0\rangle$}}
\put(22,1){\makebox(6,6)[c]{$|1\rangle$}}
\put(37,1){\makebox(6,6)[c]{$|2\rangle$}}
\put(52,1){\makebox(6,6)[c]{$|3\rangle$}}
\put(62,1){\makebox(6,6)[c]{$\cdots$}}
\put(1,7){\makebox(6,6)[c]{$|0\rangle$}}
\put(1,22){\makebox(6,6)[c]{$|1\rangle$}}
\put(1,37){\makebox(6,6)[c]{$|2\rangle$}}
\put(1,52){\makebox(6,6)[c]{$|3\rangle$}}
\put(1,62){\makebox(6,6)[c]{$\vdots$}}
\end{picture}
\vspace{-5mm}
\caption{The bases of ${\cal F}\times {\cal F}$. The black circle means 
bases giving Dirac strings, while the white one don't.}
\end{center}
\end{figure}

Then the transition ``function" (operator) is given by 
\[
\Phi_{JC}=
\left(
  \begin{array}{cc}
    a\frac{1}{\sqrt{N}} &                               \\
                        & \frac{1}{\sqrt{N}}a^{\dagger}
  \end{array}
\right)
=
\left(
  \begin{array}{cc}
    \frac{1}{\sqrt{N+1}}a &                                 \\
                          & a^{\dagger}\frac{1}{\sqrt{N+1}}
  \end{array}
\right).
\]
Therefore the system 
\begin{equation}
\label{eq:noncommutative-Hopf-bundle}
\left\{(V_{I},D_{I}), (V_{II},D_{II}), \Phi_{JC}, D_{I}\cup D_{II}\right\}
\end{equation}
is a non-commutative version of the Hopf bundle (\ref{eq:Hopf-bundle}). 
The projector in this case becomes
\begin{eqnarray}
\label{eq:quantum-projector}
P_{JC}&=&
V_{I}
\left(
  \begin{array}{cc}
    {\bf 1} &         \\
            & {\bf 0}
  \end{array}
\right)
V_{I}^{\dagger}
=
V_{II}
\left(
  \begin{array}{cc}
    {\bf 1} &         \\
            & {\bf 0}
  \end{array}
\right)
V_{II}^{\dagger}  \nonumber \\
&=&
\left\{
\begin{array}{l}
\left(
  \begin{array}{cc}
    \frac{1}{2R(N+1)} &                  \\
                      & \frac{1}{2R(N)} 
  \end{array}
\right)
\left(
  \begin{array}{cc}
      R(N+1)+\theta &  a            \\
      a^{\dagger}   &  R(N)-\theta
  \end{array}
\right)      \\
\left(
  \begin{array}{cc}
      R(N+1)+\theta &  a            \\
      a^{\dagger}   &  R(N)-\theta
  \end{array}
\right)
\left(
  \begin{array}{cc}
    \frac{1}{2R(N+1)} &                  \\
                      & \frac{1}{2R(N)} 
  \end{array}
\right).
\end{array}
\right.
\end{eqnarray}
Note that the projector $P_{JC}$ is not defined on 
${\cal F}\times \{\ket{0}\}\times \{\theta=0\}$ = ${\cal F}\times
{\cal F}\times \real - D_{I}\cup D_{II}$. 

\par \vspace{5mm}
A comment is in order. From (\ref{eq:quantum-projector}) we obtain a quantum 
version of (classical) spectral decomposition 
(a ``quantum spectral decomposition" by Suzuki \cite{TS}) 
\begin{equation}
\label{eq:quantum-spectral-decomposition}
H_{JC}=
\left(
  \begin{array}{cc}
    R(N+1) &       \\
           & R(N)
  \end{array}
\right)
P_{JC}
-
\left(
  \begin{array}{cc}
    R(N+1) &       \\
           & R(N)
  \end{array}
\right)
({\bf 1}_{2}-P_{JC}).
\end{equation}

As a bonus of the decomposition let us rederive the calculation of 
$\mbox{e}^{-igtH_{JC}}$ which has been given in \cite{MS}. 
The result is 
\begin{equation}
\mbox{e}^{-igtH_{JC}}=
\left(
  \begin{array}{cc}
   \mbox{cos}(tgR(N+1))-i\theta\frac{\mbox{sin}(tgR(N+1))}{R(N+1)}& 
   -i\frac{\mbox{sin}(tgR(N+1))}{R(N+1)}a       \\
   -i\frac{\mbox{sin}(tgR(N))}{R(N)}a^{\dagger}
   & \mbox{cos}(tgR(N))+i\theta\frac{\mbox{sin}(tgR(N))}{R(N)}
  \end{array}
\right)
\end{equation}
by making use of (\ref{eq:final-decomposition}) (or 
(\ref{eq:quantum-spectral-decomposition})). 
We leave it to the readers.

\section{Non--Commutative Veronese Mapping}

Let us make a brief review of the Veronese mapping. The map 
\[
{\fukuso}P^{1} \longrightarrow {\fukuso}P^{n}
\]
is defined as 
\[
[z_{1}:z_{2}] \longrightarrow 
\left[z_{1}^{n}:\sqrt{{}_nC_1}z_{1}^{n-1}z_{2}:\cdots:
\sqrt{{}_nC_j}z_{1}^{n-j}z_{2}^{j}:\cdots:
\sqrt{{}_nC_{n-1}}z_{1}z_{2}^{n-1}:z_{2}^{n}\right] 
\]
by making use of the homogeneous coordinate, see Appendix A.
We also have another expression of this map by using 
\[
S_{\fukuso}^{1} \longrightarrow S_{\fukuso}^{n}\ :\ 
v_{1}\equiv
\left(
  \begin{array}{c}
    z_{1} \\
    z_{2}
  \end{array}
\right)
\ \longrightarrow \
v_{n}\equiv
\left(
  \begin{array}{c}
    z_{1}^{n} \\
    \sqrt{{}_nC_1}z_{1}^{n-1}z_{2} \\
    \vdots \\
    \sqrt{{}_nC_j}z_{1}^{n-j}z_{2}^{j} \\
    \vdots \\
    \sqrt{{}_nC_{n-1}}z_{1}z_{2}^{n-1} \\
    z_{2}^{n}
  \end{array}
\right),
\quad |z_{1}|^{2}+|z_{2}|^{2}=1
\]
where 
$S_{\fukuso}^{m}=\left\{(w_{1},w_{2},\cdots,w_{m+1})^{\mbox{T}} \in 
\fukuso^{m+1}\ |\ \sum_{j=1}^{m+1}|w_{j}|^{2}=1\right\} \cong S^{2m+1}$ and 
${\fukuso}P^{m}=S_{\fukuso}^{m}/U(1)$. 
Then the Veronese mapping is also written as
\[
{\fukuso}P^{1} \longrightarrow {\fukuso}P^{n}\ :\ 
P_{1}=v_{1}v_{1}^{\dagger}
\ \longmapsto \
P_{n}=v_{n}v_{n}^{\dagger}.
\]
by using projectors.
\par \noindent
Moreover, the local map ($z\equiv z_{2}/z_{1}$) is given as 
\[
\fukuso \longrightarrow \fukuso^{n} : 
z \longrightarrow \
\left(
  \begin{array}{c}
    \sqrt{{}_nC_1}z \\
    \vdots \\
    \sqrt{{}_nC_j}z^{j} \\
    \vdots \\
    \sqrt{{}_nC_{n-1}}z^{n-1} \\
    z^{n}
  \end{array}
\right).
\]
See the following picture as a whole.

\vspace{-5mm}
\setlength{\unitlength}{1mm} 
\begin{center}
\begin{picture}(70,60)
\put(20,26){\vector(1,0){30}}
\put(20,50){\vector(1,0){30}}
\put(20, 2){\vector(1,0){30}}
\put(12,45){\vector(0,-1){15}}
\put(57,45){\vector(0,-1){15}}
\put(10,23){\makebox(6,6)[c]{${\fukuso}P^{1}$}}
\put(55,23){\makebox(6,6)[c]{${\fukuso}P^{n}$}}
\put(10,47){\makebox(6,6)[c]{$S_{\fukuso}^{1}$}}
\put(55,47){\makebox(6,6)[c]{$S_{\fukuso}^{n}$}}
\put(10,-1){\makebox(6,6)[c]{${\fukuso}$}}
\put(55,-1){\makebox(6,6)[c]{${\fukuso}^{n}$}}
\put(12, 6){\vector(0, 1){15}}
\put(57, 6){\vector(0, 1){15}}
\end{picture}
\end{center}

Next we want to consider a non--commutative version of the map. 
In the following we treat vectors as column ones. 
If we set 
\begin{equation}
{\cal A}\equiv 
\left(
  \begin{array}{c}
   X_{0} \\
   Y_{0}
  \end{array}
\right)
=
\left(
  \begin{array}{c}
    \frac{R(N+1)+\theta}{\sqrt{2R(N+1)(R(N+1)+\theta)}} \\
    \frac{1}{\sqrt{2R(N)(R(N)+\theta)}}a^{\dagger}
  \end{array}
\right)
\end{equation}
from $V_{I}$ in (\ref{eq:V-I}), then 
\[
{\cal A}^{\dagger}{\cal A}=X_{0}^{2}+Y_{0}^{\dagger}Y_{0}={\bf 1}
\quad \mbox{and}\quad 
Y_{0}X_{0}^{-1}=\frac{1}{R(N)+\theta}a^{\dagger}=Z.
\]
That is, ${\cal A}=(X_{0},Y_{0})^{T}$ is a non--commutative ``sphere" and 
$Z$ is a kind of ``stereographic projection" of the sphere. 
It is easy to see
\begin{equation}
\label{eq:easy-formula}
{\bf 1}+Z^{\dagger}Z=\frac{2R(N+1)}{R(N+1)+\theta}=X_{0}^{-2}
\Longrightarrow 
X_{0}=\left({\bf 1}+Z^{\dagger}Z\right)^{-1/2}.
\end{equation}

Here let us introduce new notations for the following. For $j\geq 0$ we set 
\begin{eqnarray}
\label{eq:def-X}
X_{-j}&=&\frac{R(N+1-j)+\theta}{\sqrt{2R(N+1-j)(R(N+1-j)+\theta)}}, \\
\label{eq:def-Y}
Y_{-j}&=&\sqrt{\frac{N-j}{N}}\frac{1}{\sqrt{2R(N-j)(R(N-j)+\theta)}}
a^{\dagger}.
\end{eqnarray}
We list some useful formulas.
\begin{equation}
\label{eq:useful-formulas}
X_{-j}^{2}+Y_{-j}^{\dagger}Y_{-j}={\bf 1}
\quad \mbox{and}\quad
Y_{-j}^{\dagger}Y_{-j}=Y_{-j+1}Y_{-j+1}^{\dagger}\quad 
\mbox{for}\quad j\geq 0.
\end{equation}

Now we are in a position to define a quantum version of the Veronese mapping 
which plays a very important role in ``classical" Mathematics.

\begin{equation}
\label{eq:Quantum-Mapping}
{\cal A}=
\left(
  \begin{array}{c}
   X_{0} \\
   Y_{0}
  \end{array}
\right)
\longrightarrow 
{\cal A}_{n}=
\left(
  \begin{array}{c}
   X_{0}^{n} \\
   \sqrt{{}_nC_1}Y_{0}X_{0}^{n-1} \\
   \vdots \\
   \sqrt{{}_nC_j}Y_{-(j-1)}Y_{-(j-2)}\cdots Y_{-1}Y_{0}X_{0}^{n-j} \\
   \vdots \\
   \sqrt{{}_nC_{n-1}}Y_{-(n-2)}Y_{-(n-3)}\cdots Y_{-1}Y_{0}X_{0} \\
   Y_{-(n-1)}Y_{-(n-2)}Y_{-(n-3)}\cdots Y_{-1}Y_{0}
  \end{array}
\right).
\end{equation}
Then it is not difficult to see 
\[
{\cal A}_{n}^{\dagger}{\cal A}_{n}=
\left(X_{0}^{2}+Y_{0}^{\dagger}Y_{0}\right)^{n}
=\left({\cal A}^{\dagger}{\cal A}\right)^{n}={\bf 1}.
\]
From this we can define the projectors which correspond to projective spaces 
like 
\begin{equation}
{\cal P}_{n}={\cal A}_{n}{\cal A}_{n}^{\dagger},\quad 
{\cal P}_{1}={\cal A}{\cal A}^{\dagger},
\end{equation}
so the map 
\begin{equation}
\label{eq:Quantum-Veronese-Mapping}
{\cal P}_{1}\ \longrightarrow \ {\cal P}_{n}
\end{equation}
is a non-commutative version of the Veronese mapping.

Next, we define a local ``coordinate" of the Veronese mapping defined above. 
\begin{eqnarray*}
{\cal A}_{n}
&=&
\left(
  \begin{array}{c}
   {\bf 1} \\
   \sqrt{{}_nC_1}Y_{0}X_{0}^{-1} \\
   \vdots \\
   \sqrt{{}_nC_j}Y_{-(j-1)}Y_{-(j-2)}\cdots Y_{-1}Y_{0}X_{0}^{-j} \\
   \vdots \\
   \sqrt{{}_nC_{n-1}}Y_{-(n-2)}Y_{-(n-3)}\cdots Y_{-1}Y_{0}X_{0}^{-(n-1)} \\
   Y_{-(n-1)}Y_{-(n-2)}Y_{-(n-3)}\cdots Y_{-1}Y_{0}X_{0}^{-n}
  \end{array}
\right)
X_{0}^{n} \\
&&{} \\
&=&\cdots \\
&&{} \\
&=&
\left(
  \begin{array}{c}
   {\bf 1} \\
   \sqrt{{}_nC_1}Y_{0}X_{0}^{-1} \\
   \vdots \\
   \sqrt{{}_nC_j}Y_{-(j-1)}X_{-(j-1)}^{-1}Y_{-(j-2)}X_{-(j-2)}^{-1}
   \cdots Y_{-1}X_{-1}^{-1}Y_{0}X_{0}^{-1} \\
   \vdots \\
   \sqrt{{}_nC_{n-1}}Y_{-(n-2)}X_{-(n-2)}^{-1}Y_{-(n-3)}X_{-(n-3)}^{-1}
   \cdots Y_{-1}X_{-1}^{-1}Y_{0}X_{0}^{-1} \\
   Y_{-(n-1)}X_{-(n-1)}^{-1}Y_{-(n-2)}X_{-(n-2)}^{-1}Y_{-(n-3)}X_{-(n-3)}^{-1}
   \cdots Y_{-1}X_{-1}^{-1}Y_{0}X_{0}^{-1}
  \end{array}
\right)
X_{0}^{n}
\end{eqnarray*}
where we have used the relation 
\[
Y_{-j}X_{-k}^{-1}=X_{-(k+1)}^{-1}Y_{-j}
\]
due to $a^{\dagger}$ in $Y_{-j}$. Moreover, by (\ref{eq:def-X}) and 
(\ref{eq:def-Y})
\[
Y_{-j}X_{-j}^{-1}=
\sqrt{\frac{N-j}{N}}\frac{1}{R(N-j)+\theta}a^{\dagger}\equiv Z_{-j}
\quad \mbox{for}\quad j\geq 0.
\]
Note that $Z_{0}=Z$. Therefore by using (\ref{eq:easy-formula}) we have 
\begin{equation}
{\cal A}_{n}=
\left(
  \begin{array}{c}
   {\bf 1} \\
   \sqrt{{}_nC_1}Z_{0} \\
   \vdots \\
   \sqrt{{}_nC_j}Z_{-(j-1)}Z_{-(j-2)}\cdots Z_{-1}Z_{0} \\
   \vdots \\
   \sqrt{{}_nC_{n-1}}Z_{-(n-2)}Z_{-(n-3)}\cdots Z_{-1}Z_{0} \\
   Z_{-(n-1)}Z_{-(n-2)}Z_{-(n-3)}\cdots Z_{-1}Z_{0}
  \end{array}
\right)
\left({\bf 1}+Z_{0}^{\dagger}Z_{0}\right)^{-n/2}.
\end{equation}

\par \noindent
Now if we define 
\begin{equation}
{\cal Z}_{n}=
\left(
  \begin{array}{c}
   \sqrt{{}_nC_1}Z_{0} \\
   \vdots \\
   \sqrt{{}_nC_j}Z_{-(j-1)}Z_{-(j-2)}\cdots Z_{-1}Z_{0} \\
   \vdots \\
   \sqrt{{}_nC_{n-1}}Z_{-(n-2)}Z_{-(n-3)}\cdots Z_{-1}Z_{0} \\
   Z_{-(n-1)}Z_{-(n-2)}Z_{-(n-3)}\cdots Z_{-1}Z_{0}
  \end{array}
\right),
\end{equation}
then 
\[
{\cal A}_{n}=
\left(
  \begin{array}{c}
   {\bf 1} \\
   {\cal Z}_{n}
  \end{array}
\right)
\left({\bf 1}+Z_{0}^{\dagger}Z_{0}\right)^{-n/2}
\]
and it is easy to show 
\[
{\bf 1}+{\cal Z}_{n}^{\dagger}{\cal Z}_{n}=
\left({\bf 1}+Z_{0}^{\dagger}Z_{0}\right)^{n},
\]
so we obtain 
\begin{eqnarray}
\label{eq:full-expression}
{\cal P}_{n}&=&{\cal A}_{n}{\cal A}_{n}^{\dagger}  \nonumber \\
&=&
\left(
  \begin{array}{c}
    {\bf 1} \\
    {\cal Z}_{n}
  \end{array}
\right)
\left({\bf 1}+{\cal Z}_{n}^{\dagger}{\cal Z}_{n}\right)^{-1}
\left({\bf 1},\ {\cal Z}_{n}^{\dagger}\right)       \nonumber \\
&=&
\left(
  \begin{array}{cc}
   \left({\bf 1}+{\cal Z}_{n}^{\dagger}{\cal Z}_{n}\right)^{-1} & 
   \left({\bf 1}+{\cal Z}_{n}^{\dagger}{\cal Z}_{n}\right)^{-1}
   {\cal Z}_{n}^{\dagger}    \\
   {\cal Z}_{n}\left({\bf 1}+{\cal Z}_{n}^{\dagger}{\cal Z}_{n}\right)^{-1} &
   {\cal Z}_{n}\left({\bf 1}+{\cal Z}_{n}^{\dagger}{\cal Z}_{n}\right)^{-1}
   {\cal Z}_{n}^{\dagger}
  \end{array}
\right)         \nonumber \\
&=&
\left(
  \begin{array}{cc}
    {\bf 1}      & -{\cal Z}_{n}^{\dagger} \\
    {\cal Z}_{n} & {\bf 1}
  \end{array}
\right)
\left(
  \begin{array}{cc}
    {\bf 1} &          \\
            & {\bf 0}
  \end{array}
\right)
\left(
  \begin{array}{cc}
    {\bf 1}       & -{\cal Z}_{n}^{\dagger} \\
     {\cal Z}_{n} & {\bf 1}
  \end{array}
\right)^{-1}.
\end{eqnarray}
This is the Oike expression in \cite{Fujii}, see also Appendix B. 

\par \vspace{3mm}
A comment is in order. Two of important properties which the classical 
Veronese mapping has are 
\begin{enumerate}
\item The Veronese mapping $\fukuso P^{1} \longrightarrow \fukuso P^{n}$ 
has the { \bf mapping degree} $n$
\item The Veronese surface (which is the image of Veronese mapping) is 
a {\bf minimal surface} in $\fukuso P^{n}$
\end{enumerate}
Since we have constructed a non--commutative version of the Veronese mapping, 
a natural question arises : What are non--commutative versions 
corresponding to 1. and 2. above ?

\par \noindent
These are very interesting problems from the view point of non--commutative 
``differential" geometry. It is worth challenging.

\section{Non--Commutative Representation Theory}

The construction of spin $j$--representation ($j \in \seisu_{\geq 0}+1/2$) 
is very well--known. Let us make a brief review within our necessity. 
For the vector space 
\[
{\cal H}_{J}=\mbox{Vect}_{\fukuso}\left\{\frac{z^{j-m}}{\sqrt{(j-m)!(j+m)!}}\ 
\vert \ 
m \in \{-j,-j+1,\cdots,j-1,j\}\right\}
\]
where $J=2j+1 (\in \futon)$, the inner product in this space is given by
\[
\left(\sum_{l=0}^{J-1}a_{l}z^{l},\ \sum_{l=0}^{J-1}b_{l}z^{l}\right)
=\sum_{l=0}^{J-1}l!(J-1-l)!a_{l}\bar{b}_{l}.
\]
For example, for $j=1/2$, $j=1$ and $j=3/2$ 
\[
{\cal H}_{2}=\mbox{Vect}_{\fukuso}\left\{z,\ 1\right\},\quad 
{\cal H}_{3}=\mbox{Vect}_{\fukuso}\left\{
\frac{z^{2}}{\sqrt{2}},z,\frac{1}{\sqrt{2}}\right\},\quad 
{\cal H}_{4}=\mbox{Vect}_{\fukuso}\left\{
\frac{z^{3}}{\sqrt{6}},\frac{z^{2}}{\sqrt{2}},\frac{z}{\sqrt{2}},
\frac{1}{\sqrt{6}}\right\}.
\]

For 
\[
A=
\left(
  \begin{array}{cc}
    \alpha & -\bar{\beta} \\
    \beta  &  \bar{\alpha}
  \end{array}
\right)\ \in SU(2)\quad \left(|\alpha|^{2}+|\beta|^{2}=1\right)
\]
the spin $j$ representation 
\[
\phi_{j} : SU(2) \longrightarrow SU(J)
\]
is defined as 
\begin{equation}
\label{eq:spin-representation}
\left(\phi_{j}(A)f\right)(z)=(-\bar{\beta}z+\bar{\alpha})^{J-1}
f\left(\frac{\alpha z+\beta}{-\bar{\beta}z+\bar{\alpha}}\right)
\end{equation}
where $f \in {\cal H}_{J}$. It is easy to obtain $\phi_{j}(A)$ for $j=1/2$, 
$j=1$ and $j=3/2$.

Namely, the spin 1/2 representation is 
\begin{equation}
\label{eq:spin-1/2}
\phi_{1/2}(A)=
\left(
  \begin{array}{cc}
    \alpha & -\bar{\beta} \\
    \beta  &  \bar{\alpha}
  \end{array}
\right)=A,
\end{equation}
the spin 1 representation is 
\begin{equation}
\label{eq:spin-1}
\phi_{1}(A)=
\left(
  \begin{array}{ccc}
    \alpha^{2} & -\sqrt{2}\alpha\bar{\beta} & \bar{\beta}^{2} \\
    \sqrt{2}\alpha\beta & |\alpha|^{2}-|\beta|^{2} & 
    -\sqrt{2}\bar{\alpha}\bar{\beta} \\
    \beta^{2} & \sqrt{2}\bar{\alpha}\beta & \bar{\alpha}^{2}
  \end{array}
\right),
\end{equation}
and the spin 3/2 representation is 
\begin{equation}
\label{eq:spin-3/2}
\phi_{3/2}(A)=
\left(
  \begin{array}{cccc}
    \alpha^{3} & -\sqrt{3}\alpha^{2}\bar{\beta} & 
    \sqrt{3}\alpha\bar{\beta}^{2} & -\bar{\beta}^{3} \\
    \sqrt{3}\alpha^{2}\beta & (|\alpha|^{2}-2|\beta|^{2})\alpha & 
    -(2|\alpha|^{2}-|\beta|^{2})\bar{\beta} & 
    \sqrt{3}\bar{\alpha}\bar{\beta}^{2} \\
    \sqrt{3}\alpha\beta^{2} & (2|\alpha|^{2}-|\beta|^{2})\beta & 
    (|\alpha|^{2}-2|\beta|^{2})\bar{\alpha} & 
    -\sqrt{3}\bar{\alpha}^{2}\bar{\beta} \\
    \beta^{3} & \sqrt{3}\bar{\alpha}\beta^{2} & 
    \sqrt{3}\bar{\alpha}^{2}\beta & \bar{\alpha}^{3}
  \end{array}
\right).
\end{equation}

\par \vspace{5mm}
Next we want to consider a non--commutative version of the spin 
representation. However, since such a theory has not been known as far as 
we know we must look for mappings corresponding to $\phi_{1}(A)$ and 
$\phi_{3/2}(A)$ by (many) trial and error, see Appendix C.

If we set 
\[
V\equiv V_{I}=
\left(
  \begin{array}{cc}
    X_{0} & -Y_{0}^{\dagger} \\
    Y_{0} &  X_{-1}
  \end{array}
\right)\ :\ \mbox{unitary}
\]
from (\ref{eq:V-I}), then the corresponding map for $\phi_{1}(A)$ is 
\begin{equation}
\label{eq:non-spin-1}
\Phi_{1}(V)=
\left(
 \begin{array}{ccc}
  X_{0}^{2}& -\sqrt{2}X_{0}Y_{0}^{\dagger}& Y_{0}^{\dagger}Y_{-1}^{\dagger} \\
  \sqrt{2}Y_{0}X_{0}& X_{-1}^{2}-Y_{-1}^{\dagger}Y_{-1}& 
  -\sqrt{2}X_{-1}Y_{-1}^{\dagger} \\
  Y_{-1}Y_{0}& \sqrt{2}Y_{-1}X_{-1}& X_{-2}^{2}
  \end{array}
\right)
\end{equation}             
and the corresponding map for $\phi_{3/2}(A)$ is 
\begin{eqnarray}
\label{eq:non-spin-3/2}
\Phi_{3/2}(V)&=&
\left(
 \begin{array}{cccc}
  X_{0}^{3}& 
  -\sqrt{3}X_{0}^{2}Y_{0}^{\dagger}& 
  \sqrt{3}X_{0}Y_{0}^{\dagger}Y_{-1}^{\dagger}& 
  -Y_{0}^{\dagger}Y_{-1}^{\dagger}Y_{-2}^{\dagger} \\
  \sqrt{3}Y_{0}X_{0}^{2}& 
  X_{-1}\left(X_{-1}^{2}-2Y_{-1}^{\dagger}Y_{-1}\right)&
  -\left(2X_{-1}^{2}-Y_{-1}^{\dagger}Y_{-1}\right)Y_{-1}^{\dagger}&
  \sqrt{3}X_{-1}Y_{-1}^{\dagger}Y_{-2}^{\dagger} \\
  \sqrt{3}Y_{-1}Y_{0}X_{0}&
  Y_{-1}\left(2X_{-1}^{2}-Y_{-1}^{\dagger}Y_{-1}\right)&
  X_{-2}\left(X_{-2}^{2}-2Y_{-2}^{\dagger}Y_{-2}\right)&
  -\sqrt{3}X_{-2}^{2}Y_{-2}^{\dagger} \\
  Y_{-2}Y_{-1}Y_{0}&
  \sqrt{3}Y_{-2}Y_{-1}X_{-1}&
  \sqrt{3}Y_{-2}X_{-2}^{2}&
  X_{-3}^{3}
  \end{array}
\right).      \nonumber \\
&&{}
\end{eqnarray}
To check the unitarity of $\Phi_{1}(V)$ and $\Phi_{3/2}(V)$ is long but 
straightforward. 

For $j\geq 2$ we could not find a general method like 
(\ref{eq:spin-representation}) which determines $\Phi_{j}(V)$. 
However, we know only that the first column of $\Phi_{j}(V)$ is just 
${\cal A}_{2j}$ in (\ref{eq:Quantum-Mapping}).,
\[
\Phi_{j}(V)=\left({\cal A}_{2j},*,\cdots,*\right)\ :\ \mbox{unitary}
\]
and 
\[
\Phi_{j}(V)
\left(
 \begin{array}{ccccc}
 {\bf 1} &         &       &       &         \\
         & {\bf 0} &       &       &         \\
         &         & \cdot &       &         \\
         &         &       & \cdot &         \\
         &         &       &       & {\bf 0} \\
  \end{array}
\right)
\Phi_{j}(V)^{\dagger}
=
{\cal A}_{2j}{\cal A}_{2j}^{\dagger}
=
{\cal P}_{2j}.
\]

\par \noindent
We leave finding a general method to the readers as a challenging problem.

\section{Discussion}

In this paper we derived a non--commutative version of the Berry model 
from the Jaynes--Cummings model in quantum optics and constructed a 
non--commutative version of the Hopf bundle. 

The bundle has a kind of Dirac strings. However, they appear in only states 
containing the ground one 
(${\cal F}\times \{\ket{0}\} \cup \{\ket{0}\}\times {\cal F} \subset 
{\cal F}\times {\cal F}$) 
and don't appear in excited states, which is very interesting. 

In general, a non-commutative version of classical field theory is of course 
not unique. If our model is a ``correct" one, then this paper give an example 
that classical singularities like Dirac strings are not universal in some 
non--commutative model. 
As to general case with higher spins which are not easy see \cite{TS}.

More generally, it is probable that a singularity (singularities) in some 
classical model is (are) removed in the process of non--commutativization. 

Moreover, based on this model a non--commutative version of the Veronese 
mapping or spin representation was constructed. 
The results in the paper will become a starting point to construct 
fruitful non--commutative geometry.

Last, we would like to make a comment. To develop a ``quantum" mathematics 
we need a rigorous method to treat an analysis or a geometry on infinite 
dimensional spaces like Fock space. 
In quantum field theories physicists have given some (interesting) methods, 
while they are more or less formal from the mathematical point of view. 
It is a rigorous method which we need. 
As a trial \cite{Asada} is recommended.

\vspace{5mm}
\noindent{\em Acknowledgment.}\\
The author wishes to thank Akira Asada, Yoshinori Machida, Ryu Sasaki and 
Tatsuo Suzuki for their helpful comments and suggestions.

\par \vspace{10mm}
\begin{center}
 \begin{Large}
   {\bf Appendix}
 \end{Large}
\end{center}

\vspace{5mm} \noindent
\begin{Large}
{\bf A\ \ Classical Theory of Projective Spaces}
\end{Large}

\par \vspace{3mm} \noindent
Complex projective spaces are typical examples of symmetric spaces and are 
very tractable, so they are used to construct several examples in both 
physics and mathematics. 

\par \noindent
We make a review of complex projective spaces within our necessity, see for 
example \cite{MN}, \cite{Fujii}, \cite{FKSF}. 

For $n \in \futon$ the complex projective space ${\fukuso}P^{n}$ is defined 
as follows : 
For \mbox{\boldmath $\zeta$}, \mbox{\boldmath $\mu$} $\in 
{\fukuso}^{n+1}-\{{\bf 0}\}$  \mbox{\boldmath $\zeta$} is equivalent to 
\mbox{\boldmath $\mu$} (\mbox{\boldmath $\zeta$} $\sim$ \mbox{\boldmath $\mu$}) if and only if 
\mbox{\boldmath $\zeta$} = $\lambda$\mbox{\boldmath $\mu$} for 
$\lambda \in \fukuso - \{0 \}$. 
We show the equivalent relation class as [\mbox{\boldmath $\zeta$}] and set 
${\fukuso}P^{n} \equiv {\fukuso}^{n+1}-\{{\bf 0}\} / \sim $. 
For   
\mbox{\boldmath $\zeta$} = $({\zeta}_{0}, {\zeta}_{1}, \cdots, {\zeta}_{n})$ 
we write usually as [\mbox{\boldmath $\zeta$}] = $[{\zeta}_{0}: {\zeta}_{1}:  
\cdots : {\zeta}_{n}]$. Then it is well--known that ${\fukuso}P^{n}$ has 
$n+1$ local charts, namely
\begin{equation}
  {\fukuso}P^{n} = \bigcup_{j=0}^{n} U_{j}\ ,  \quad 
    U_{j} = \{ [{\zeta}_{0}: \cdots : {\zeta}_{j}: \cdots : {\zeta}_{n}]\ |\  
          {\zeta}_{j} \ne 0 \}.
\end{equation} 
Since
\[
  ({\zeta}_{0}, \cdots , {\zeta}_{j}, \cdots , {\zeta}_{n}) =  
  {\zeta}_{j}\left(\frac{{\zeta}_{0}}{{\zeta}_{j}}, \cdots, 
  \frac{{\zeta}_{j-1}}{{\zeta}_{j}}, 1, \frac{{\zeta}_{j+1}}{{\zeta}_{j}}, 
  \cdots, \frac{{\zeta}_{n}}{{\zeta}_{j}}\right),
\]
we have the local coordinate on $U_{j}$ 
\begin{equation}
  \left(\frac{{\zeta}_{0}}{{\zeta}_{j}}, \cdots, 
  \frac{{\zeta}_{j-1}}{{\zeta}_{j}}, \frac{{\zeta}_{j+1}}{{\zeta}_{j}}, 
  \cdots, \frac{{\zeta}_{n}}{{\zeta}_{j}}\right). 
\end{equation}

However the above definition of ${\fukuso}P^{n}$ is not tractable, so we use 
the well--known expression by projections
\begin{equation}
 {\fukuso}P^{n} \cong G_{1}({\fukuso}^{N+1}) = 
     \{P \in M(N+1; \fukuso)\ |\ P^{2} = P,\ P^{\dagger} = P \ \mbox{and}\ 
       \mbox{tr}P = 1 \}
\end{equation}
and the correspondence 
\begin{equation}
 \label{eq:correspondence}
  [{\zeta}_{0}: {\zeta}_{1}: \cdots : {\zeta}_{n}] \Longleftrightarrow 
  \frac{1}{\zetta{{\zeta}_{0}}^2 + \zetta{{\zeta}_{1}}^2 + \cdots + 
          \zetta{{\zeta}_{n}}^2 }
  \left(
     \begin{array}{ccccc} 
         \zetta{{\zeta}_{0}}^2& {\zeta}_{0}{\bar {\zeta}_{1}}& 
         \cdot& \cdot& {\zeta}_{0}{\bar {\zeta}_{n}}  \\
         {\zeta}_{1}{\bar {\zeta}_{0}} & \zetta{{\zeta}_{1}}^2&  
         \cdot& \cdot& {\zeta}_{1}{\bar {\zeta}_{n}}  \\
         \cdot& \cdot& & & \cdot  \\
         \cdot& \cdot& & & \cdot  \\
         {\zeta}_{n}{\bar {\zeta}_{0}}& {\zeta}_{n}{\bar {\zeta}_{1}}& 
         \cdot& \cdot& \zetta{{\zeta}_{n}}^2     
     \end{array}
  \right) \equiv P\ .
\end{equation}
If we set 
\begin{equation}
  \ket{\mbox{\boldmath $\zeta$}}=
 \frac{1}{\sqrt{\sum_{j=0}^{n} \zetta{\zeta_{j}}^2} }
  \left(
     \begin{array}{c}
        {\zeta}_{0} \\
        {\zeta}_{1} \\
         \cdot  \\
         \cdot  \\
        {\zeta}_{n} 
     \end{array} 
  \right)\ , 
\end{equation}
then we can write the right hand side of (\ref{eq:correspondence}) as 
\begin{equation}
 \label{eq:projection}
  P = \ket{\mbox{\boldmath $\zeta$}}\bra{\mbox{\boldmath $\zeta$}} \quad 
  \mbox{and} \quad 
    \braket{\mbox{\boldmath $\zeta$}}{\mbox{\boldmath $\zeta$}} = 1.
\end{equation}
For example on $U_{0}$ 
\[
  \left(z_{1}, z_{2}, \cdots, z_{n} \right) = 
  \left(\frac{{\zeta}_{1}}{{\zeta}_{0}},\frac{{\zeta}_{2}}{{\zeta}_{0}}, 
  \cdots, 
  \frac{{\zeta}_{n}}{{\zeta}_{0}}\right) ,  
\]
we have 
\begin{eqnarray}
  P(z_{1}, \cdots, z_{n}) &=& 
  \frac{1}{1 + \sum_{j=1}^{n} \zetta{z_{j}}^2}
     \left(
         \begin{array}{ccccc}
             1& {\bar z_{1}}& \cdot& \cdot& {\bar z_{n}}  \\
             z_{1}& \zetta{z_{1}}^2& \cdot& \cdot& z_{1}{\bar z_{n}} \\
             \cdot& \cdot& & & \cdot \\
             \cdot& \cdot& & & \cdot \\
             z_{n}& z_{n}{\bar z_{1}}& \cdot& \cdot& \zetta{z_{n}}^2
         \end{array}
     \right)  \nonumber \\
   &=& \ket{\left(z_{1}, z_{2}, \cdots, z_{n}\right)}
       \bra{\left(z_{1}, z_{2}, \cdots, z_{n}\right)}\ ,
\end{eqnarray}
where 
\begin{equation}
  \ket{\left(z_{1}, z_{2}, \cdots, z_{n} \right)} = 
  \frac{1}{\sqrt{1 + \sum_{j=1}^{n} \zetta{z_{j}}^2}}
     \left(
         \begin{array}{c}
             1 \\
             z_{1} \\
             \cdot \\
             \cdot \\
             z_{n} 
         \end{array}
     \right).  \nonumber \\
\end{equation}

To be clearer, let us give a detailed description for the case of $n$ = $1$ 
and $2$.

\par \noindent
(a) {\bf $n = 1$} : 
\begin{eqnarray}
 \label{eq:cp1-1}
  P(z)&=&\frac{1}{1+\zetta{z}^2}
     \left(
         \begin{array}{cc}
             1& {\bar z} \\
             z& \zetta{z}^2 
         \end{array}
     \right)  
   = \ket{z}\bra{z}, \nonumber  \\
  &&\mbox{where}\ \ket{z}=\frac{1}{\sqrt{1+\zetta{z}^2}}
     \left(
         \begin{array}{c}
             1 \\
             z 
         \end{array}
     \right), 
  \quad z=\frac{\zeta_{1}}{\zeta_{0}}, \quad \mbox{on}\ U_{0}\ ,  \\
 \label{eq:cp1-2}
  P(w)&=&\frac{1}{\zetta{w}^2+1}
     \left(
         \begin{array}{cc}
             \zetta{w}^2 & w \\
             {\bar w}& 1
         \end{array}
     \right)  
   = \ket{w}\bra{w},  \nonumber  \\
  &&\mbox{where}\ \ket{w}=\frac{1}{\sqrt{\zetta{w}^2+1}}
     \left(
         \begin{array}{c}
             w \\
             1
         \end{array}
     \right), 
  \quad w=\frac{\zeta_{0}}{\zeta_{1}}, \quad \mbox{on}\ U_{1}\ . 
\end{eqnarray}

\vspace{5mm}
\par \noindent
(b) {\bf $n = 2$} : 
\begin{eqnarray}
 \label{eq:cp2-1}
  P(z_{1},z_{2})&=&\frac{1}{1+\zetta{z_{1}}^2+\zetta{z_{2}}^2}
     \left(
         \begin{array}{ccc}
             1& {\bar z_{1}}& {\bar z_{2}} \\
             z_{1}& \zetta{z_{1}}^2& z_{1}{\bar z_{2}} \\
             z_{2}& z_{2}{\bar z_{1}}& \zetta{z_{2}}^2 
         \end{array}
     \right)  
   = \ket{(z_{1},z_{2})}\bra{(z_{1},z_{2})}, \nonumber  \\
  \mbox{where} 
  &&\ket{(z_{1},z_{2})}=\frac{1}{\sqrt{1+\zetta{z_{1}}^2+\zetta{z_{2}}^2}}
     \left(
         \begin{array}{c}
             1 \\
             z_{1} \\
             z_{2} 
         \end{array}
     \right), 
\quad (z_{1},z_{2})=\left(\frac{\zeta_{1}}{\zeta_{0}},
   \frac{\zeta_{2}}{\zeta_{0}} \right)  \quad \mbox{on}\ U_{0}\ ,  \\
 \label{eq:cp2-2}
  P(w_{1},w_{2})&=&\frac{1}{\zetta{w_{1}}^2+1+\zetta{w_{2}}^2}
     \left(
         \begin{array}{ccc}
             \zetta{w_{1}}^2& w_{1}& w_{1}{\bar w_{2}} \\
             {\bar w_{1}}& 1& {\bar w_{2}} \\
             w_{2}{\bar w_{1}}& w_{2}& \zetta{w_{2}}^2 
         \end{array}
     \right)  
   = \ket{(w_{1},w_{2})}\bra{(w_{1},w_{2})}, \nonumber \\
  \mbox{where} 
  &&\ket{(w_{1},w_{2})}=\frac{1}{\sqrt{\zetta{w_{1}}^2+1+\zetta{w_{2}}^2}}
     \left(
         \begin{array}{c}
             w_{1} \\
              1  \\
             w_{2} 
         \end{array}
     \right), 
\quad  (w_{1},w_{2})=\left(\frac{\zeta_{0}}{\zeta_{1}},
   \frac{\zeta_{2}}{\zeta_{1}} \right)\ \  \mbox{on}\ U_{1}\ , \\
 \label{eq:cp2-3}
  P(v_{1},v_{2})&=&\frac{1}{\zetta{v_{1}}^2+\zetta{v_{2}}^2+1}
     \left(
         \begin{array}{ccc}
             \zetta{v_{1}}^2& v_{1}{\bar v_{2}}& v_{1} \\
             v_{2}{\bar v_{1}}& \zetta{v_{2}}^2& v_{2} \\
             {\bar v_{1}}& {\bar v_{2}}& 1 
         \end{array}
     \right)  
   = \ket{(v_{1},v_{2})}\bra{(v_{1},v_{2})}, \nonumber \\
  \mbox{where} 
  &&\ket{(v_{1},v_{2})}=\frac{1}{\sqrt{\zetta{v_{1}}^2+\zetta{v_{2}}^2+1}}
     \left(
         \begin{array}{c}
             v_{1} \\
             v_{2}  \\
              1 
         \end{array}
     \right), 
\quad (v_{1},v_{2})=\left(\frac{\zeta_{0}}{\zeta_{2}},
   \frac{\zeta_{1}}{\zeta_{2}} \right)  \quad \mbox{on}\ U_{2}\ . 
\end{eqnarray}

\vspace{5mm} \noindent
\begin{Large}
{\bf B\ \ Local Coordinate of the Projector}
\end{Large}
\par \vspace{3mm} \noindent
We give a proof to the last formula in (\ref{eq:full-expression}). 

By making use of the expression by Oike in \cite{Fujii} (we don't 
repeat it here) 
\begin{equation}
\label{eq:oike-expression-1}
{\cal P}({\cal Z})=
\left(
  \begin{array}{cc}
    {\bf 1}  & -{\cal Z}^{\dagger} \\
    {\cal Z} & {\bf 1}
  \end{array}
\right)
\left(
  \begin{array}{cc}
    {\bf 1} &          \\
            & {\bf 0}
  \end{array}
\right)
\left(
  \begin{array}{cc}
    {\bf 1}  & -{\cal Z}^{\dagger} \\
    {\cal Z} & {\bf 1}
  \end{array}
\right)^{-1}
\end{equation}
where ${\cal Z}$ is some operator on the Fock space ${\cal F}$. 
Let us rewrite this into more useful form. From the simple relation 
\[
\left(
  \begin{array}{cc}
    {\bf 1}   & {\cal Z}^{\dagger} \\
    -{\cal Z} & {\bf 1}
  \end{array}
\right)
\left(
  \begin{array}{cc}
    {\bf 1}  & -{\cal Z}^{\dagger} \\
    {\cal Z} & {\bf 1}
  \end{array}
\right)
=
\left(
  \begin{array}{cc}
    {\bf 1}+{\cal Z}^{\dagger}{\cal Z}  &               \\
                  & {\bf 1}+{\cal Z}{\cal Z}^{\dagger}
  \end{array}
\right)
\]
we have 
\[
\left(
  \begin{array}{cc}
    {\bf 1}  & -{\cal Z}^{\dagger} \\
    {\cal Z} & {\bf 1}
  \end{array}
\right)^{-1}
=
\left(
  \begin{array}{cc}
    ({\bf 1}+{\cal Z}^{\dagger}{\cal Z})^{-1}  &          \\
             & ({\bf 1}+{\cal Z}{\cal Z}^{\dagger})^{-1}
  \end{array}
\right)
\left(
  \begin{array}{cc}
    {\bf 1}  & {\cal Z}^{\dagger} \\
   -{\cal Z} & {\bf 1}
  \end{array}
\right).
\]
Inserting this into (\ref{eq:oike-expression-1}) and some calculation leads to 
\begin{equation}
\label{eq:oike-expression-2}
{\cal P}({\cal Z})=
\left(
  \begin{array}{cc}
  ({\bf 1}+{\cal Z}^{\dagger}{\cal Z})^{-1} & 
  ({\bf 1}+{\cal Z}^{\dagger}{\cal Z})^{-1}{\cal Z}^{\dagger}        \\
  {\cal Z}({\bf 1}+{\cal Z}^{\dagger}{\cal Z})^{-1} & 
  {\cal Z}({\bf 1}+{\cal Z}^{\dagger}{\cal Z})^{-1}{\cal Z}^{\dagger}
  \end{array}
\right).
\end{equation}

\par \vspace{3mm}
Comparing (\ref{eq:oike-expression-2}) with (\ref{eq:quantum-projector}) 
we obtain the ``local coordinate" 
\begin{equation}
\label{eq:quantum local coordinate}
{\cal Z}=\frac{1}{R(N)+\theta}a^{\dagger}=a^{\dagger}\frac{1}{R(N+1)+\theta}
\end{equation}
where $R(N)=\sqrt{N+\theta^{2}}$. ${\cal Z}$ obtained by ``stereographic 
projection" is a kind of complex coordinate. 

Now if we take a classical limit $a \longrightarrow x-iy$, 
$a^{\dagger} \longrightarrow x+iy$ and $\theta=z$ then 
\begin{equation}
Z_{c}=\frac{x+iy}{r+z}
\end{equation}
where $r=\sqrt{x^{2}+y^{2}+z^{2}}$. This is nothing but a well--known one for 
(\ref{eq:projector}).

\vspace{5mm} \noindent
\begin{Large}
{\bf C\ \ Difficulty of Tensor Decomposition}
\end{Large}
\par \vspace{3mm} \noindent
We point out a difficulty in obtaining the formula (\ref{eq:non-spin-1}) 
or (\ref{eq:non-spin-3/2}) by decomposing tensor products of $V$. 

To obtain the formula (\ref{eq:spin-1}) there is another method which 
uses a decomposition of the tensor product $A\otimes A$. Let us introduce. 
For 
\[
A=
\left(
  \begin{array}{cc}
    \alpha & -\bar{\beta} \\
    \beta  &  \bar{\alpha}
  \end{array}
\right)\ \in SU(2)
\]
we have 
\[
A\otimes A
=
\left(
  \begin{array}{cccc}
  \alpha^{2} & -\alpha\bar{\beta} & -\alpha\bar{\beta} & \bar{\beta}^{2} \\
  \alpha\beta & |\alpha|^{2} & -|\beta|^{2} & -\bar{\alpha}\bar{\beta}   \\
  \alpha\beta & -|\beta|^{2} & |\alpha|^{2} & -\bar{\alpha}\bar{\beta}   \\
  \beta^{2} & \bar{\alpha}\beta & \bar{\alpha}\beta & \bar{\alpha}^{2}
  \end{array}
\right).
\]
For the matrix $T$ coming from the Clebsch--Gordan decomposition 
\[
T=
\left(
  \begin{array}{cccc}
   0 & 1 & 0 & 0                                     \\
   \frac{1}{\sqrt{2}}  & 0 & \frac{1}{\sqrt{2}} & 0  \\
   -\frac{1}{\sqrt{2}} & 0 & \frac{1}{\sqrt{2}} & 0  \\
   0 & 0 & 0 & 1
  \end{array}
\right)
\]
it is easy to see 
\begin{equation}
T^{\dagger}(A\otimes A)T
=
\left(
  \begin{array}{cccc}
  |\alpha|^{2}+|\beta|^{2} &   &   &                             \\
    & \alpha^{2} & -\sqrt{2}\alpha\bar{\beta} & \bar{\beta}^{2}  \\
    & \sqrt{2}\alpha\beta & |\alpha|^{2}-|\beta|^{2} & 
      -\sqrt{2}\bar{\alpha}\bar{\beta}                           \\
    & \beta^{2} & \sqrt{2}\bar{\alpha}\beta & \bar{\alpha}^{2}
  \end{array}
\right)
=
\left(
  \begin{array}{cc}
   1 &             \\
     & \phi_{1}(A)
  \end{array}
\right)
\end{equation}
where we have used $|\alpha|^{2}+|\beta|^{2}=1$. This means a well--known 
decomposition
\[
\frac{1}{2}\otimes \frac{1}{2}=0\oplus 1.
\]

Let us take an analogy. For 
\[
V=
\left(
  \begin{array}{cc}
    X_{0} & -Y_{0}^{\dagger} \\
    Y_{0} &  X_{-1}
  \end{array}
\right)
\]
we have 
\[
V\otimes V
=
\left(
  \begin{array}{cccc}
   X_{0}^{2} & -X_{0}Y_{0}^{\dagger} & -Y_{0}^{\dagger}X_{0} & 
   Y_{0}^{\dagger}Y_{0}^{\dagger} \\
   X_{0}Y_{0} & X_{0}X_{-1} & -Y_{0}^{\dagger}Y_{0} & 
   -Y_{0}^{\dagger}X_{-1} \\
   Y_{0}X_{0} & -Y_{0}Y_{0}^{\dagger} & X_{-1}X_{0} & 
   -X_{-1}Y_{0}^{\dagger} \\
   Y_{0}Y_{0} & Y_{0}X_{-1} & X_{-1}Y_{0} & X_{-1}^{2}
  \end{array}
\right).
\]
However, the analogy breaks down at this stage because of 
the non--commutativity 
\begin{equation}
T^{\dagger}(V\otimes V)T
\ne
\left(
  \begin{array}{cc}
   {\bf 1} &              \\
           & \Phi_{1}(V)
  \end{array}
\right)
\end{equation}
for (\ref{eq:non-spin-1}). We leave it to the readers. 
There is no (well--known) direct method to obtain $\Phi_{1}(V)$ at the 
current time.

\par \vspace{5mm}
Last, let us make a comment. 
For the matrix $T$ coming from the Clebsch--Gordan decomposition (see 
\cite{papers})
\[
T=
\left(
  \begin{array}{cccccccc}
    0 & 0 & 0 & 0 & 1 & 0 & 0 & 0 \\
    \frac{1}{\sqrt{2}} & 0 & \frac{1}{\sqrt{6}} & 0 & 0 & 
    \frac{1}{\sqrt{3}} & 0 & 0 \\
    -\frac{1}{\sqrt{2}} & 0 & \frac{1}{\sqrt{6}} & 0 & 0 & 
    \frac{1}{\sqrt{3}} & 0 & 0 \\ 
    0 & 0 & 0 & \frac{\sqrt{2}}{\sqrt{3}} & 0 & 0 & \frac{1}{\sqrt{3}} & 0 \\
    0 & 0 & -\frac{\sqrt{2}}{\sqrt{3}} & 0 & 0 & \frac{1}{\sqrt{3}} & 0 & 0 \\
    0 & \frac{1}{\sqrt{2}} & 0 & -\frac{1}{\sqrt{6}} & 0 & 0 & 
    \frac{1}{\sqrt{3}} & 0 \\
    0 & -\frac{1}{\sqrt{2}} & 0 & -\frac{1}{\sqrt{6}} & 0 & 0 & 
    \frac{1}{\sqrt{3}} & 0 \\
    0 & 0 & 0 & 0 & 0 & 0 & 0 & 1
  \end{array}
\right)
\]
it is not difficult to see 
\begin{eqnarray}
&&T^{\dagger}(A\otimes A\otimes A)T   \nonumber \\
=
&&
\left(
  \begin{array}{cccccccc}
  \alpha & -\bar{\beta}  &   &   &   &   &   &       \\
  \beta  &  \bar{\alpha} &   &   &   &   &   &       \\
    &   & \alpha & -\bar{\beta}  &   &   &   &       \\
    &   & \beta  &  \bar{\alpha} &   &   &   &       \\
    &   &   &   & \alpha^{3} & -\sqrt{3}\alpha^{2}\bar{\beta} & 
    \sqrt{3}\alpha\bar{\beta}^{2} & -\bar{\beta}^{3} \\
    &   &   &   & \sqrt{3}\alpha^{2}\beta & (|\alpha|^{2}-2|\beta|^{2})\alpha 
    & -(2|\alpha|^{2}-|\beta|^{2})\bar{\beta} & 
    \sqrt{3}\bar{\alpha}\bar{\beta}^{2}              \\
    &   &   &   & \sqrt{3}\alpha\beta^{2} & (2|\alpha|^{2}-|\beta|^{2})\beta 
    & (|\alpha|^{2}-2|\beta|^{2})\bar{\alpha} & 
    -\sqrt{3}\bar{\alpha}^{2}\bar{\beta}             \\
    &   &   &   & \beta^{3} & \sqrt{3}\bar{\alpha}\beta^{2} & 
    \sqrt{3}\bar{\alpha}^{2}\beta & \bar{\alpha}^{3}
  \end{array}
\right)           \nonumber \\
=
&&
\left(
  \begin{array}{ccc}
  \phi_{1/2}(A) &     &     \\
      & \phi_{1/2}(A) &     \\
      &    & \phi_{3/2}(A)
  \end{array}
\right).
\end{eqnarray}
This means a well--known decomposition
\[
\frac{1}{2}\otimes \frac{1}{2}\otimes \frac{1}{2}=
\left(0\oplus 1\right)\otimes \frac{1}{2}=
\left(0\otimes \frac{1}{2}\right)\oplus
\left(1\otimes \frac{1}{2}\right)=
\frac{1}{2}\oplus \frac{1}{2}\oplus \frac{3}{2}.
\]
%


\end{document}